\documentclass{osa-article}

\graphicspath{{./fig/}}
\usepackage{subfigure}

\journal{oe}

\articletype{Research Article}

\begin{document}

\title{Asymptotic analysis of V-BLAST MIMO for coherent optical wireless communications in Gamma-Gamma turbulence}

\author{Yiming Li,\authormark{1} Chao Gao,\authormark{1} Mark S. Leeson,\authormark{2} and Xiaofeng Li\authormark{1,*}}

\address{\authormark{1}School of Astronautics and Aeronautics, University of Electronic Science and Technology of China, Chengdu, Sichuan, 611731 China\\
\authormark{2}School of Engineering, University of Warwick, Coventry, CV4 7AL UK}

\email{\authormark{*}lxf3203433@uestc.edu.cn} 



\begin{abstract}
This paper investigates the asymptotic BER performance of coherent optical wireless communication systems in Gamma-Gamma turbulence when applying the V-BLAST MIMO scheme. A new method is proposed to quantify the performance of the system and mathematical solutions for asymptotic BER performance are derived. Counterintuitive results are shown since the diversity gain of the V-BLAST MIMO system is equal to the number of the receivers. As a consequence, it is shown that when applying the V-BLAST MIMO scheme, the symbol rate per transmission can be equal to the number of transmitters with some cost to diversity gain. This means that we can simultaneously exploit the spatial multiplexing and diversity properties of the MIMO system to achieve a higher data rate than existing schemes in a channel that displays severe turbulence and moderate attenuation.
\end{abstract}

\section{Introduction}
\label{sec:1}
Optical wireless communication (OWC) systems have received a great deal of attention in research and commerce recently. This type of system has a very high bandwidth, which is unlicensed and therefore allows extremely high data rate transmission (of the order of Gigabits per second). With a very narrow laser beam, the OWC system also provides a very high reuse factor, inherent security and robustness to electromagnetic interference \cite{Khalighi2014p2231}.

Intensity modulation and direct detection (IM/DD) with on/off keying (OOK) are the most established approaches to OWC systems owing to their low realization complexity. However, recent advances in digital signal processing (DSP) have significantly increased the practicality of, and reduced the difficulty of, coherent system implementation. Considering their excellent background light rejection, very high power efficiency and enhanced frequency selectivity \cite{Chan2006p4750}, coherent optical systems have thus drawn great attention from both the scholarly and the industrial communities. 

When a laser propagates freely in an unconstrained medium, it will suffer from channel turbulence, which may cause severe signal degradation. Multiple-Input Multiple-Output (MIMO) technology is very popular as a method to increase data rates as well as mitigate fading effects in radio frequency (RF) communication systems \cite{Paulraj2004p198}. However, in OWC communication, most MIMO systems are proposed to reduce the turbulence induced fading by employing Repetition Coding (RC) in IM/DD systems \cite{Bayaki2009p3415, Bhatnagar2016p2158}, with the optimality of RC for IM/DD having been recently proved by Zhang et al. \cite{Zhang2016p846} and the capacity of the IM/DD systems has been discussed by Wang et al. \cite{Wang2016Channel}.

Unlike DD schemes, coherent OWC systems have an extra degree of freedom to transmit signals by exploiting the phase information as well as the amplitude information. Therefore, higher diversity gain can be achieved and high-order quadrature amplitude modulation (QAM) constellations can be created in coherent systems. However, the inherent phase noise in such systems means that RC will suffer from severe performance degradation and thus a new spatial coding method is needed \cite{niu2012coherent}. Niu et al. \cite{Niu2014p7900217} have demonstrated the availability of an Alamouti-type Space Time Block Code (A-STBC) scheme in coherent MIMO OWC communication systems, which exploits full diversity gain but without exploiting the multiplexing property. Moreover, Zhang et al. have demonstrated the possibility of applying A-STBC on independent orbital angular momentum (OAM) modes to further increase the transmitted bit rate \cite{Zhang2017Atmospheric}.

On the other hand, the multiplexing gain, which can be exploited to achieve a much higher data rate, is another crucial property of the MIMO system. However, the lack of phase information and non-negative property precludes the application of the multiplexing property in IM/DD systems \cite{Wang2016Channel}. Fortunately, coherent systems, which can simultaneously exploit both phase and amplitude information, are suitable for exploiting the multiplexing property. Zhang et al. demonstrated the possibility of applying the Vertical-Bell Laboratories Layered Space-Time (V-BLAST) code to fully exploit the multiplexing property and achieve a much higher data rate in coherent MIMO OWC systems \cite{Zhang2017Atmospheric}. Moreover, Zhou et al.'s research on the channel capacity of coherent MIMO OWC channels also implies that the multiplexing gain can be exploited in coherent MIMO OWC systems \cite{Zhou2018Atmospheric}. 

However, references \cite{Zhang2017Atmospheric} and \cite{Zhou2018Atmospheric} are based on numerical simulation results rather than analytical derivations. To the best of our knowledge, no research has carried out a rigorous performance derivation of the V-BLAST scheme in coherent MIMO OWC systems. 
Although comparative analysis can be carried out in RF communications \cite{Zheng2003p1073, Grant1998p1038, Grant2000p1788} , the properties of Gaussian matrices that can be used in Rayleigh channels are no longer available in turbulent OWC channels. Therefore, the derivation of the BER performance in V-BLAST MIMO OWC systems is significantly more challenging than in the RF regime.
Considering that analytical solutions will deliver enhanced insight into the coherent V-BLAST MIMO OWC system, we propose a new method to analyze the asymptotic BER performance of such a system in this paper. Rigorous derivation has been carried out to analyze the asymptotic BER performance in addition to the diversity gain in the V-BLAST MIMO OWC system. It is shown from the results that such a system can improve the transmitted symbol rate with some costs to diversity gain. We can also conclude that it is possible for us to transmit at a higher bit rate than existing A-STBC and RC schemes in a channel that displays severe turbulence and moderate attenuation.


The rest of the paper is organized as follows: Section~\ref{sec:2} outlines the system and channel models. In Section~\ref{sec:3} the asymptotic BER performance and diversity gain of the coherent MIMO OWC V-BLAST system in Gamma-Gamma channels are derived. The analytical results are compared with simulations in Section~\ref{sec:4} followed by conclusions in Section~\ref{sec:5}.

\section{Preliminaries}
\label{sec:2}

\subsection{System model}
\label{sec:2.1}

\begin{figure}[htb]
  \centering
  \includegraphics[width=4.5in]{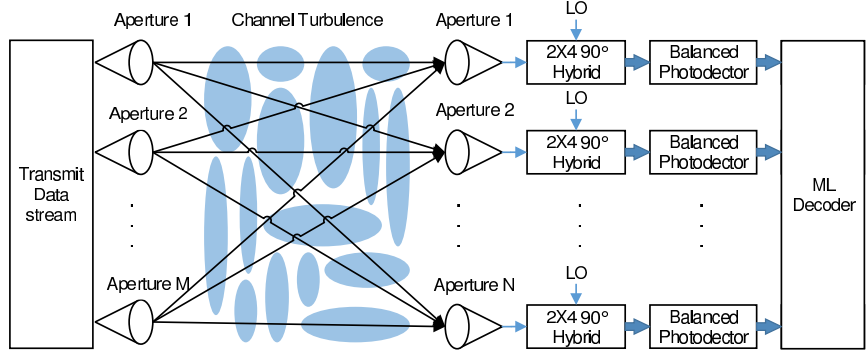}
  \caption{V-BLAST MIMO coherent OWC system.}
  \label{fig:1}
  \end{figure}

We consider a V-BLAST MIMO coherent OWC system with \(M\) transmit and \(N\) receive apertures. The block diagram of this system is shown in Fig.~\ref{fig:1}.
Given the inherent property of the V-BLAST regime, \(M\) independent coherent phase shift keying (PSK) or QAM signals are simultaneously transmitted through \(M\) different transmit apertures. At the receiver, the power of the local oscillator (LO) is equally divided between the \(2 \times 4\) \(90^\circ\) hybrids. Moreover, it is assumed that the receiver implements Maximum-Likelihood Detection (MLD) and has perfect channel state information at the receiver (CSIR) such as fading gain and phase information, which is a common situation in the slow fading regime. We also assume that the channel response is flat and that the dynamic range of the system is effectively infinite to simplify the analysis. 

The received electric field at the \({n^{th}}\), \( n \in \{1,2,\cdots,N\}\) receive aperture from the  \({m^{th}}\), \(m \in \{1,2,\cdots,M\}\) transmit aperture, is given by:
  \begin{equation}\label{equ:1}
  {e_{nm}}\left( t \right) = \sqrt {{P_{nm}}}  \cdot {A_{s,m}}\exp \left( {j\left( {\omega t + {\phi _{nm}} + {\phi _{s,m}}} \right)} \right),
  \end{equation}
where \({{P_{nm}}}\) is the received power from the \({m^{th}}\) transmitted signal (which is subject to optical scintillation), \(\omega\) denotes the optical carrier frequency of the signal laser, \({\phi _{nm}}\) can be modeled as a Wiener process which represents the overall phase noise from the \({m^{th}}\) transmitter to the input of the \({n^{th}}\) coherent receiver, \(A_{s,m}\) is the encoded amplitude information, and \({\phi _{s,m}}\) is the encoded phase information.

The electric field of the LO can be expressed as:
  \begin{equation}\label{equ:2}
  {e_{LO}}\left( t \right) = \sqrt {{P_{LO}}} \exp \left( {j\left( {{\omega _{LO}} t + {\phi _{LO}}} \right)} \right),
  \end{equation}
where  \({{P_{LO}}}\) is the power of the LO, \({\omega _{LO}}\) denotes the optical carrier frequency of the LO, and \({\phi _{LO}}\) can be also modeled as a Wiener process which represents the phase noise from the LO.

After mixing the received signal of the \({n^{th}}\) aperture with the LO by a \(2 \times 4\) \(90^\circ\) hybrid and detection by balanced photodetectors \cite{seimetz2006options}, the received electrical signal at the \({n^{th}}\) receive aperture, following optical-to-electrical (O-E) conversion, is:
  \begin{equation}\label{equ:3}
  \left\{ {\begin{array}{*{20}{c}}
{{i_1}\left( t \right) = \frac{1}{4} \cdot {R_{oe}}{{\left |{{e_{LO}}\left( t \right) + \sum\limits_{m = 1}^M {{e_{nm}}\left( t \right)} } \right |}^2}}\\
{{i_2}\left( t \right) = \frac{1}{4} \cdot {R_{oe}}{{\left| {j{e_{LO}}\left( t \right) + \sum\limits_{m = 1}^M {{e_{nm}}\left( t \right)} } \right|}^2}}\\
{{i_3}\left( t \right) = \frac{1}{4} \cdot {R_{oe}}{{\left| { - {e_{LO}}\left( t \right) + \sum\limits_{m = 1}^M {{e_{nm}}\left( t \right)} } \right|}^2}}\\
{{i_4}\left( t \right) = \frac{1}{4} \cdot {R_{oe}}{{\left| { - j{e_{LO}}\left( t \right) + \sum\limits_{m = 1}^M {{e_{nm}}\left( t \right)} } \right|}^2}}
\end{array}} \right.,
  \end{equation}
  where \(R_{oe}\) denotes the photodiode responsitivity, and \({i_1}\left( t \right)\), \({i_2}\left( t \right)\), \({i_3}\left( t \right)\), and \({i_4}\left( t \right)\) denote the \(0^\circ \), \(90^\circ \), \(180^\circ \), and \(270^\circ \) channels respectively.
  
  Considering the in-phase channel:
  \begin{equation}\label{equ:4}
  {i_I}\left( t \right) = {i_1}\left( t \right) - {i_3}\left( t \right).
  \end{equation}
  
  Substituting Eq.~\eqref{equ:3} into Eq.~\eqref{equ:4}, assuming that the LO phase is zero and that the LO frequency is filtered out, we can simplify Eq.~\eqref{equ:4} to:
  \begin{equation}\label{equ:5}
  {i_I}\left( t \right) = \sum\limits_{m = 1}^M {{{\eta _{nm}}}  \cdot {A_{s,m}} \cos \left( {{\phi _{nm}} - {\phi _{LO}} + {\phi _{s,m}}} \right)},
  \end{equation}
where \(\eta_{nm} = {R_{oe}\sqrt {{P_{nm}}{P_{LO}}}}\).

Similarly, the quadrature channel signal can be written as:
  \begin{equation}\label{equ:6}
  {i_Q}\left( t \right) = \sum\limits_{m = 1}^M {{{\eta _{nm}}}  \cdot {A_{s,m}} \sin \left( {{\phi _{nm}} - {\phi _{LO}} + {\phi _{s,m}}} \right)}.
  \end{equation}

Considering Eqs.~\eqref{equ:5}~and~\eqref{equ:6}, the received signal of the \(n^{th}\) receive aperture can be written as:
  \begin{equation}\label{equ:7}
  \begin{aligned}
{y_n}\left( t \right) = {i_I}\left( t \right) + {i_Q}\left( t \right) + {n_n}\left( t \right) = \sum\limits_{m = 1}^M {{\eta _{nm}} \cdot {A_{s,m}} \exp \left( {j\left( {\Delta {\phi _{nm}} + {\phi _{s,m}}} \right)} \right)}  + {n_n}\left( t \right),
  \end{aligned}
  \end{equation}
where \(\Delta {\phi _{nm}}=\phi _{nm} - \phi _{LO}\) can be modeled as a Wiener process which is uniformly distributed between 0 and \(2\pi\) in independent observations. The term \({n_n(t)}\) denotes the noise at the \({n^{th}}\) receive aperture which is approximately circular symmetric additive white Gaussian noise (AWGN) when the power of the local oscillator is sufficiently large, making the receiver shot noise limited.

Moreover, the channel between the multiple transmit and receive apertures can be described by the channel gain matrix:
  \begin{equation}\label{equ:8}
  {\bf{H}} = \left[ {\begin{array}{*{20}{c}}
{h_{11}} & {h_{12}} & \cdots & {h_{1M}}\\
{h_{21}} & {h_{22}} & \cdots & {h_{2M}}\\
\vdots   & \vdots   & \ddots & \vdots  \\
{h_{N1}} & {h_{N2}} & \cdots & {h_{NM}}
\end{array}} \right],
  \end{equation}
where \(h_{nm} = \eta _{nm} e^{j \Delta \phi _{nm}}\). When \(h_{nm}\) is normalized, \({\left\| {{h_{nm}}} \right\|^2} = {I_{nm}}\) will obey the normalized Gamma-Gamma distribution given by Eq.~\eqref{equ:12}.
  
Therefore, we can write Eq.~\eqref{equ:7} in matrix form for brevity as:
  \begin{equation}\label{equ:9}
  {\bf{y}} = {\bf{Hs}} + {\bf{n}},
  \end{equation}
where the transmitted signal vector is given by \({\bf{s}} = {\left[ {{{A_{s,1}}{e^{j\Delta {\phi _{s,1}}}}},{{A_{s,2}}{e^{j\Delta {\phi _{s,2}}}}}, \cdots ,{{A_{s,M}}{e^{j\Delta {\phi _{s,M}}}}}} \right]^T}\), the received signal vector by \({\bf{y}} ~= ~{\left[ {{y_1}\left( t \right),{y_2}\left( t \right), \cdots ,{y_N}\left( t \right)} \right]^T}\), and the noise vector by \({\bf{n}} ~= ~\left[ n_1(t),n_2(t), \cdots , n_N(t)\right] ^T\). When \(\bf{s}\) and \(\bf{H}\) are normalized, each component of \(\bf{n}\) should obey the distribution \({n_i}\left( t \right)\sim CN\left( {0,\frac{1}{{SNR}}} \right)\). The average SNR of the optical receiver can be written as \cite{Derr1989p127}:
  \begin{equation}\label{equ:10}
  SNR = \frac{{2{T_b}R_{oe}{\bar I_s}}}{{h\nu }},
  \end{equation}
where \(T_b\) denotes the symbol period, \(\bar I_s\) the average irradiance of the channel, \(h\) is Planck's constant and \(\nu\) is the frequency of the lightwave.

\subsection{Maximum-likelihood detection (MLD)}
\label{sec:2.2}

When the probability of each transmitted signal is independent and remains identically distributed, MLD is the optimum approach as it minimizes the probability of error. In this paper, we consider MLD to decode the received signal and achieve the optimal V-BLAST performance. Because the additive noise of the system is approximately AWGN, the MLD decision criterion can be written as:
  \begin{equation}\label{equ:11}
  {\bf{\hat s}} = \mathop {\arg \min }\limits_{\bf{s}} {\left\| {{\bf{y}} - {\bf{Hs}}} \right\|^2},
  \end{equation}
where \(\bf{\hat s}\) is the estimated symbol vector, and \(\left\|  \cdot  \right\|\) is the Euclidean norm on the complex space.

\subsection{Channel model}
\label{sec:2.3}

To evaluate the performance of FSO systems, it is necessary to use an appropriate model to describe the fading characteristics induced by atmospheric turbulence. The lognormal distribution is often used to model weak turbulence conditions whereas the K-distribution is used to model strong turbulence. Moreover, the exponentiated Weibull distribution was proposed to model the turbulence under aperture averaging conditions \cite{barrios2012exponentiated}. On the other hand, generalized models such as the Gamma-Gamma distribution has also been proposed to describe scintillation over arbitrary turbulence conditions \cite{andrews2005laser}. In recent years, the \(\mathcal{M}\)~distribution has also been proposed as a generalization of most existing distribution models by using a summation of Gamma-Gamma functions \cite{jurado2011unifying}. We accept that for given scenarios it will be possible to select a distribution that may deliver a better fit. However, we wish to present general insight into system performance that is as broad as possible. Considering the trade-off between accuracy and mathematical simplicity, we use the Gamma-Gamma model to describe the generalized turbulence in this paper. When applying this probability model, the distribution of the normalized intensity scintillation can be expressed as \cite{andrews2005laser}:
  \begin{equation}\label{equ:12}
  {f_{{I}}}\left( {{I}} \right) = \frac{2}{{\Gamma (\alpha )\Gamma (\beta )}}{(\alpha \beta )^{\frac{{\alpha  + \beta }}{2}}}{I}^{\frac{{\alpha  + \beta }}{2} - 1}{K_{\alpha  - \beta }}\left( {2\sqrt {\alpha \beta {I}} } \right),
  \end{equation}
where \({K_\nu }\left(  \cdot  \right)\) is a \(\nu^{th} \) order modified Bessel function of the second kind, and \(\alpha \) and \(\beta \) are the effective numbers of the small-scale and large-scale eddies in the scattering environment. The two scales are implemented mathematically by inserting heuristic spatial-frequency filters into weak-turbulence integrals for the scintillation index.

On the other hand, the atmospheric attenuation is deterministic and may be constant for many hours, which can be described by the exponential Beer-Lambert law \cite{naboulsi2005propagation}. Moreover, the attenuation and the atmospheric turbulence will introduce independent influences to the system and can be discussed separately \cite{touati2016effects,Bhatnagar2016p2158}. Therefore, this paper is focused on the influence of turbulence induced fading rather than the atmospheric attenuation.

\section{Asymptotic analysis}
\label{sec:3}

In RF communications, the BER performance of a system can be easily understood when the SNR is high. Therefore, we follow a similar path here for optical communications and present the asymptotic analysis of the V-BLAST MIMO scheme BER performance.

In this section, we consider a V-BLAST MIMO system with an arbitrary modulation constellation. To analyze the asymptotic BER performance, we first discuss some preliminaries for calculating the Pairwise Error Probability (PEP), which is shown in Sec.~\ref{sec:3.1}. Secondly, we consider the equivalent integral area for the PEP in a MISO system in Sec.~\ref{sec:3.2} under the constraint of high SNR. Thirdly, we generalize the equivalent integral area to a MIMO system in Sec.~\ref{sec:3.3}. Fourthly, by applying the conclusions in Sec.~\ref{sec:3.1} and Sec.~\ref{sec:3.3}, we determine the crucial PEP of more than one symbol error in Sec.~\ref{sec:3.4}. We then derive the diversity gain as well as the asymptotic BER performance of the generalized V-BLAST system by using the result in Sec.~\ref{sec:3.4} and applying previous conclusions drawn from \cite{john2008digital,Niu2012p6515}.

\subsection{MLD pairwise error probability (PEP)}
\label{sec:3.1}

To analyze the asymptotic performance of the BER, we need to calculate the PEP first. If we transmit a symbol vector \(\bf{s}\) but \(\bf{s'}\) is detected, the PEP can be written as:
  \begin{equation}\label{equ:13}
  P\left( {{\bf{s}} \to {\bf{s'}}} \right) = P\left\{ {\left\| {{\bf{y}} - {\bf{Hs}}} \right\| > \left\| {{\bf{y}} - {\bf{Hs'}}} \right\|} \right\}.
  \end{equation}

By substituting Eq.~\eqref{equ:9} into Eq.~\eqref{equ:13}, an equivalent form of Eq.~\eqref{equ:13} is given as:
  \begin{equation}\label{equ:14}
  P\left( {{\bf{s}} \to {\bf{s'}}} \right) = P\left\{ {\left\| {\bf{n}} \right\| > \left\| {{\bf{H}}\left( {{\bf{s}} - {\bf{s'}}} \right) + {\bf{n}}} \right\|} \right\},
  \end{equation}
where \(\left\| {\bf{H}}\left( {{\bf{s}} - {\bf{s'}}} \right) \right\| \) can be regarded as the Euclidean distance in the N-dimensional space (N represents the number of the receivers). Therefore, the noise vector \(\bf{n}\) can be projected onto the direction of the vector \( {\bf{H}}\left( {{\bf{s}} - {\bf{s'}}} \right) \) and the PEP can be calculated as:
  \begin{equation}\label{equ:15}
  P\left( {{\bf{s}} \to {\bf{s'}}} \right) = P\left\{ {n > \frac{1}{2}\left\| {{\bf{H}}\left( {{\bf{s}} - {\bf{s'}}} \right)} \right\|} \right\} = P\left\{ {n > \frac{1}{2}\left\| {{\bf{H}}\Delta {\bf{s}}} \right\|} \right\},
  \end{equation}
where \(\Delta {\bf{s}} = {\bf{s}} - {\bf{s'}}\), and \(n \sim N\left( {0,\frac{1}{{2SNR}}} \right)\) which is the projected equivalent noise of the system.

\subsection{Probability distribution of the MISO system in equivalent integral area}
\label{sec:3.2}

As a precursor to the calculation of the the PEP of a MIMO system with arbitrary receivers, we first analyze the case of \(N=1\) as a special situation when \(\bf{H}\) reduces to a row vector. Therefore Eq.~\eqref{equ:15} becomes simply:
  \begin{equation}\label{equ:16}
  P\left( {{\bf{s}} \to {\bf{s'}}} \right) = P\left\{ {n > \frac{1}{2}\left| {\sum\limits_{1 \le j \le k} {{h_{1{m_j}}}\Delta {s_{{m_j}}}} } \right|} \right\},
  \end{equation}
where \(k\) is the total number of non-zero elements of \(\Delta\bf{s}\), and \(m_j\) is the corresponding subscript of the \(j^{th}\) non-zero element which satisfies \(1 \le {m_1} <  \cdots  < {m_k} \le M\).

\begin{figure}[htb]
  \centering
  \includegraphics[width=3in]{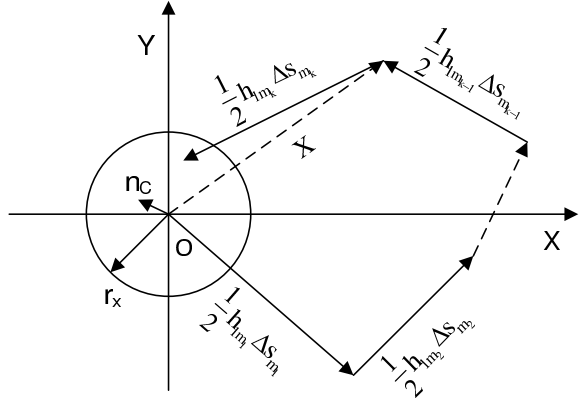}
  \caption{Effective integral area of the MISO system in the complex plane.}
  \label{fig:2}
  \end{figure}

When \(k=1\), the problem reduces to calculating the symbol error rate (SER) in a SISO system, and we can find the asymptotic analysis in Niu et al.'s previous work \cite{Niu2012p6515}. So the problem reduces to calculating Eq.~\eqref{equ:16} when \(k \ge 2\). On the other hand, the additive noise can be expanded in the complex plane and will obey the complex Gaussian distribution \({n_C} \sim CN\left( {0,\frac{1}{{SNR}}} \right)\), which is shown in Fig.~\ref{fig:2}.

To calculate the PEP of such a system, we define an auxiliary parameter \(r_x = SN{R^{ - {n_r}}}\), where \({n_r} \in \left( {0,\frac{1}{2}} \right)\), to simplify the calculation. There are two important reasons for this definition. Firstly, we can conclude that when \(SNR \to \infty \), \(r_x\) will be infinitesimal, and this property will simplify our analysis. Secondly, considering that \(\sigma _{{n_c}}^2 = \frac{1}{{SNR}}\), we can additionally see that \(\sigma _{{n_c}}^2/{r_x^2}\) will also be infinitesimal. This implies that the noise will fall into a circle of radius \(r_x\) with probability 1. Therefore we can calculate the PEP with the constraint of \({\frac{1}{2}\left| {\sum\limits_{1 \le j \le k} {{h_{1{m_j}}}\Delta {s_{{m_j}}}} } \right|} < r_x\), which will be equal to the real PEP when \(SNR \to \infty \).

As is shown in the Appendix, for arbitrary \(r<r_x\), the cumulative distribution function (cdf) of \({\frac{1}{2}\left| {\sum\limits_{1 \le j \le k} {{h_{1{m_j}}}\Delta {s_{{m_j}}}} } \right|} < r\) can be written as:
\begin{equation}\label{equ:17}
  {F_1}\left( r \right) = {C_r}{r^2},
  \end{equation}
and the probability density function (pdf) is thus:
  \begin{equation}\label{equ:18}
{f_1}\left( r \right) = \frac{{{\rm{d}}{F_1}\left( r \right)}}{{{\rm{d}}r}} = 2{C_r}r.
  \end{equation}
The form of \(C_r\) is:
  \begin{equation}\label{equ:19}
  {C_r} = \int\limits_{\bf{I}} {\frac{4}{{{{\left| {\Delta {s_{{m_k}}}} \right|}^2}}}{f_I}\left( {\frac{{4{{\left| X \right|}^2}}}{{{{\left| {\Delta {s_{{m_k}}}} \right|}^2}}}} \right){f_{\bf{I}}}\left( {\bf{I}} \right){\rm{d}}{\bf{I}}},
  \end{equation}
where \(X = \frac{1}{2}{\sum\limits_{j = 1}^{k - 1} {{h_{1{m_j}}}\Delta {s_{{m_j}}}} } \), \({\bf{I}} = \left[ {{I_{1{m_1}}},{I_{1{m_2}}}, \ldots ,{I_{1{m_{k - 1}}}}} \right]\), and \({f_{\bf{I}}}\left( {\bf{I}} \right) = \sum\limits_{j = 1}^{k - 1} {{f_I}\left( {{I_{{1m_j}}}} \right)} \). Eq.~\eqref{equ:19} can be numerically calculated by most mathematics software packages such as MATLAB. 

Furthermore, when we constrain the system to two transmitters and binary phase shift keying (BPSK) modulation, we have \(\Delta s_{m_k} = 2\) and \(\left | X \right| ^2 = I_{11}\). Therefore, Eq.~\eqref{equ:19} reduces to:
  \begin{equation}\label{equ:20}
  {C_r} = \frac{1}{\pi }\int_0^{ + \infty } {f_I^2\left( {{I_{11}}} \right){\rm{d}}{I_{11}}}.
  \end{equation}
Substituting Eq.~\eqref{equ:12} into Eq.~\eqref{equ:20}, replacing \(I_{11}\) by \(x^2\), and then using the Eq.~(6.576.4) in \cite{jeffrey2007table}, we obtain the closed form of \(C_r\) as:
  \begin{equation}\label{equ:21}
  {C_r} = \frac{{\Gamma \left( {\alpha  + \beta  - 1} \right)\Gamma \left( {2\beta  - 1} \right)\Gamma \left( {2\alpha  - 1} \right)}}{{{\Gamma ^2}\left( \alpha  \right){\Gamma ^2}\left( \beta  \right)\Gamma \left( {\alpha  + \beta  - 1/2} \right)}} \cdot {2^{3 - 2\alpha  - 2\beta }}\sqrt \pi   \cdot \alpha \beta.
  \end{equation}

\subsection{Probability distribution of the MIMO system in equivalent integral area}
\label{sec:3.3}

Considering Eq.~\eqref{equ:15}, we need to calculate the cdf of the system with \(N\) receivers:
  \begin{equation}\label{equ:22}
  {F_N}\left( r \right) = P\left\{ {\frac{1}{2}\left\| {{\bf{H}}\Delta {\bf{s}}} \right\| < r} \right\}.
  \end{equation}

We can conclude that the cdf and pdf of such systems can be expressed respectively as:
  \begin{equation}\label{equ:23}
  {F_N}\left( r \right) = \frac{{C_r^N}}{{N!}}{r^{2N}},
  \end{equation}
and
  \begin{equation}\label{equ:24}
  {f_N}\left( r \right) = \frac{{{\rm{d}}{F_N}\left( r \right)}}{{{\rm{d}}r}} = \frac{{2C_r^N}}{{\left( {N - 1} \right)!}}{r^{2N - 1}}.
  \end{equation}

Eqs.~\eqref{equ:23}~and~\eqref{equ:24} can be proved by mathematical induction as below:

If \(N=1\), it is clear that Eqs.~\eqref{equ:17}~and~\eqref{equ:18} satisfy Eqs.~\eqref{equ:23}~and~\eqref{equ:24}.

If \(N \ge 2\), we assume the conclusion is true for \(N-1\), then
  \begin{equation}\label{equ:25}
  \begin{aligned}
  {F_N}\left( r \right) &=
  \int_0^r {P\left\{ {{{\left| {\frac{1}{2}{{\bf{h}}_N}\Delta {\bf{s}}} \right|}^2} < {r^2} - r_0^2} \right\}} {f_{N - 1}}\left( {{r_0}} \right){\rm{d}}{r_0}\\
 &= \int_0^r {{C_r}\left( {{r^2} - r_0^2} \right)}  \cdot \frac{{2C_r^{N - 1}}}{{\left( {N - 2} \right)!}}{r^{2N - 3}}{\rm{d}}{r_0}\\
 &= \frac{{C_r^N}}{{N!}}{r^{2N}},
  \end{aligned}
  \end{equation}
where \({\bf{h}}_i\) represents the \(i^{th}\) row vector of \(\bf{H}\). Thus Eqs.~\eqref{equ:23}~and~\eqref{equ:24} are proved by extension to \(N\).

\subsection{PEP of the MIMO system}
\label{sec:3.4}

As previously stated, \(\sigma _{{n_c}}^2/{r_x^2}\) will be infinitesimal as \(SNR \to \infty \) implying that the noise vector will fall into a circle of radius \(r_x\) with probability 1. Therefore we can calculate the PEP with the constraint of \(\frac{1}{2}\left\| {{\bf{H}}\Delta {\bf{s}}} \right\| < {r_x}\), which will be equal to the true PEP when \(SNR \to \infty \). So the PEP can be expressed as:
  \begin{equation}\label{equ:26}
  \mathop {\lim }\limits_{SNR \to \infty } P\left( {\bf{s}} \to {\bf{s}}' \right) = \int_0^{r_x} {Q\left( {t\sqrt {2 \cdot SNR} } \right) \cdot {f_N}\left( t \right){\rm{d}}t}.
  \end{equation}

where \(Q \left( x \right)\) is the Gaussian Q-function, \(Q\left( x \right) < \frac{1}{{x\sqrt {2\pi } }}{e^{ - \frac{{{x^2}}}{2}}}\), which decays much faster than any rational function. Therefore, we have: 
  \begin{equation}\label{equ:27}
  \mathop {\lim }\limits_{SNR \to \infty } \int_{r_x}^{ + \infty } {Q\left( {t\sqrt {2 \cdot SNR} } \right) \cdot {f_N}\left( t \right){\rm{d}}t}  = 0.
  \end{equation}

Considering Eqs.~\eqref{equ:26}~and~\eqref{equ:27}, we can calculate the PEP as below:
  \begin{equation}\label{equ:28}
  \begin{aligned}
\mathop {\lim }\limits_{SNR \to \infty } P\left( {\bf{s}} \to {\bf{s}}' \right) & = \int_0^{ + \infty } {Q\left( {t\sqrt {2 \cdot SNR} } \right) \cdot {f_N}\left( t \right){\rm{d}}t} \\
& = \int_0^{ + \infty } {\frac{1}{2}{\rm{erfc}}\left( {\frac{x}{{\sqrt 2 }}} \right) \cdot {x^{2N - 1}} \cdot \frac{{2C_r^N}}{{\left( {N - 1} \right)!}} \cdot \frac{1}{{{{\left( {2 \cdot SNR} \right)}^N}}}{\rm{d}}x} \\
& = \frac{{C_r^N}}{{2\sqrt \pi   \cdot N!}} \cdot \Gamma \left( {N + \frac{1}{2}} \right) \cdot \frac{1}{{{{\left( {SNR} \right)}^N}}}.
  \end{aligned}
  \end{equation}

\subsection{BER performance of the MIMO system}
\label{sec:3.5}

Considering the inequalities below \cite{john2008digital}:
  \begin{equation}\label{equ:29}
  \frac{1}{{{q^N}}}\mathop {\max }\limits_{{\bf{s}} \ne {\bf{s'}}} \left\{ {{\sigma _{{\bf{ss'}}}}P\left( {{\bf{s}} \to {\bf{s'}}} \right)} \right\} \le {P_e} \le \frac{1}{{{q^N}}}\sum\limits_{{\bf{s}} \ne {\bf{s'}}} {{\sigma _{{\bf{ss'}}}}P\left( {{\bf{s}} \to {\bf{s'}}} \right)},
  \end{equation}
where \(q\) is the number of points in the constellation, \(N\) is the number of receivers, and \(\sigma _{\bf{ss'}}\) is the BER when \(\bf{s}\) is transmitted but \(\bf{s'}\) is detected. According to the conclusions in Niu et al.'s previous work \cite{Niu2012p6515}, we can conclude that when \(k=1\), \(P\left( {{\bf{s}} \to {\bf{s'}}} \right) \) is proportional to \(SN{R^{ - N\min \left\{ {\alpha ,\beta } \right\}}}\) when \(SNR \to \infty\), which is the higher order infinitesimal to the \(k \ge 2\) case given by Eq.~\eqref{equ:28}. Therefore, using Eq.~\eqref{equ:28} and Eq.~\eqref{equ:29} and taking logarithms, we will arrive at:
  \begin{equation}\label{equ:30}
  \left\{ \begin{aligned}
\mathop {\lim }\limits_{SNR \to \infty } \frac{{\log {P_e}}}{{\log SNR}} &\ge \mathop {\lim }\limits_{SNR \to \infty } \frac{{\log \left( {\frac{1}{{{q^N}}}\mathop {\max }\limits_{{\bf{s}} \ne {\bf{s'}}} \left\{ {{\sigma _{{\bf{ss'}}}}P\left( {{\bf{s}} \to {\bf{s'}}} \right)} \right\}} \right)}}{{\log SNR}} =  - N\\
\mathop {\lim }\limits_{SNR \to \infty } \frac{{\log {P_e}}}{{\log SNR}} &\le \mathop {\lim }\limits_{SNR \to \infty } \frac{{\log \left( {\frac{1}{{{q^N}}}\sum\limits_{{\bf{s}} \ne {\bf{s'}}} {{\sigma _{{\bf{ss'}}}}P\left( {{\bf{s}} \to {\bf{s'}}} \right)} } \right)}}{{\log SNR}} =  - N
\end{aligned} \right..
  \end{equation}
That is to say, the diversity gain of the generalized MIMO system with an arbitrary modulation constellation can be obtained as:
  \begin{equation}\label{equ:31}
  {d^*} =  - \mathop {\lim }\limits_{SNR \to \infty } \frac{{\log {P_e}}}{{\log SNR}} = N,
  \end{equation}
and there exists a constant \(C_x\) which satisfies the equation:
  \begin{equation}\label{equ:32}
  \mathop {\lim }\limits_{SNR \to \infty } {P_e} = {C_x} \cdot SN{R^{ - N}}.
  \end{equation}
  
We can conclude from Eqs.~\eqref{equ:31}~and~\eqref{equ:32} that the diversity gain of such a system will be \(N\) rather than \(N \cdot \min \left\{ {\alpha ,\beta } \right\}\), which is a common situation in SIMO systems. That is to say, transmitting at a symbol rate of \(M\) symbols per transmission will come at the expense of diversity gain.

Furthermore, when we constraint the system to 2 transmitters and BPSK modulation, we can restrict the transmitted data to \(\left[ 1,1 \right] ^T\) without loss of generality by exploiting the circular symmetric nature of the channel parameter \(h_{nm}\). Therefore, Eq.~\eqref{equ:29} is reduced to:
  \begin{equation}\label{equ:33}
  P\left( {11 \to 00} \right) \le {P_e} \le P\left( {11 \to 00} \right) + \frac{1}{2}P\left( {11 \to 10} \right) + \frac{1}{2}P\left( {11 \to 01} \right).
  \end{equation}
and \(P\left( {11 \to 10} \right)\) and \(P\left( {11 \to 01} \right)\) are both higher order infinitesimal to \(P\left( {11 \to 00} \right)\) when \(SNR \to \infty\). So we can conclude that both the left side and the right side of Eq.~\eqref{equ:33} will converge to \(P\left( {11 \to 00} \right)\) when \(SNR \to \infty\). Therefore, by applying Eq.~\eqref{equ:28}, Eq.~\eqref{equ:32} can be written as:
  \begin{equation}\label{equ:34}
  \mathop {\lim }\limits_{SNR \to \infty } {P_e} = \frac{{C_r^N}}{{2\sqrt \pi   \cdot N!}} \cdot \Gamma \left( {N + \frac{1}{2}} \right) \cdot \frac{1}{{{{\left( {SNR} \right)}^N}}},
  \end{equation}
where \(C_r\) is expressed in Eq.~\eqref{equ:21}.

\section{Numerical results}
\label{sec:4}
In this section, we provide results based on the analytical expressions derived for the V-BLAST MIMO coherent scheme above. These are compared with the outcomes of \(10^9\) independent Monte-Carlo simulations.

  \begin{figure}[htb]
  \centering
  \includegraphics[width=5in]{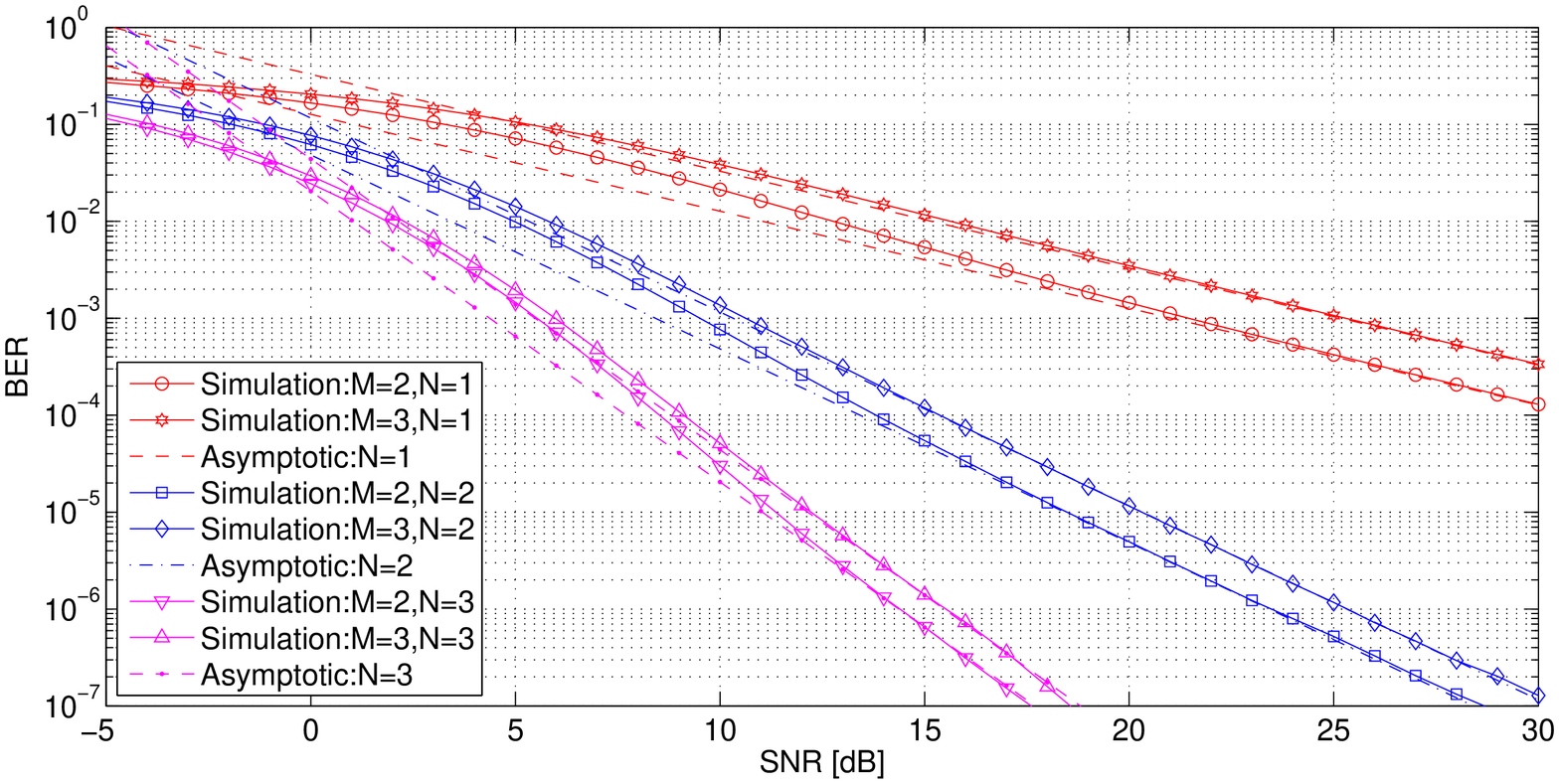}
  \caption{BER of coherent V-BLAST MIMO systems: \(M=2,3\), \(N=1,2,3\), \(\alpha=4\), \(\beta=2\).}
  \label{fig:3}
  \end{figure}
In  Fig.~\ref{fig:3}, we consider a system with \(M=2,3\), \(N=1,2,3\). The Gamma-Gamma turbulence parameters are \(\alpha=4\), \(\beta=2\) which represent moderate turbulence. As a result of the behavior of Eqs.~\eqref{equ:32}~and~\eqref{equ:34}, the numerical BER curves converge to the analytical asymptotic lines, and the diversity gain of the V-BLAST MIMO system with two transmitters is equal to \(N\) rather than \(N \cdot \min \left\{ {\alpha ,\beta } \right\}\), which means that the higher transmitted symbol rate is at the cost of diversity gain. However, considering the situation of severe turbulence, in which  \(\beta\) is just slightly larger than unity, the cost will be acceptable in a channel with moderate attenuation. On the other hand, compared to the system with two transmitters, There is an obvious SNR penalty of between 1 dB and 4 dB in the high SNR area when \(M=3\). This is because the larger number of transmitters will introduce additional interference. 

\begin{figure*}[htb]
  \centering
  \subfigure[\(\alpha=4\), \(\beta=1, 1.5, 2, 2.5\).]{
    \centering
    \includegraphics[width=2.53in]{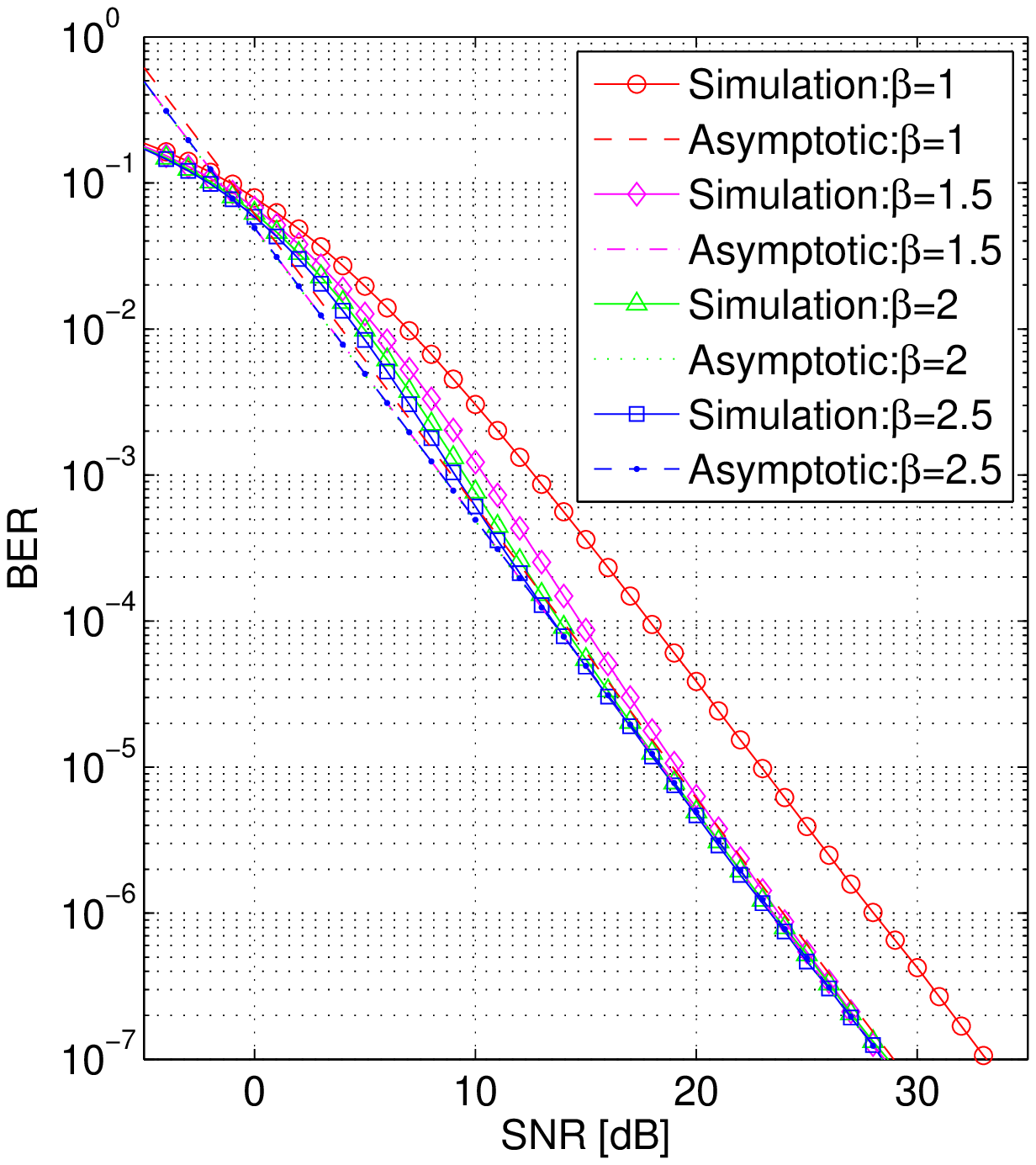}
  }
  \subfigure[\(\alpha=2, 3, 4\), \(\beta=1.5\).]{
    \centering
    \includegraphics[width=2.53in]{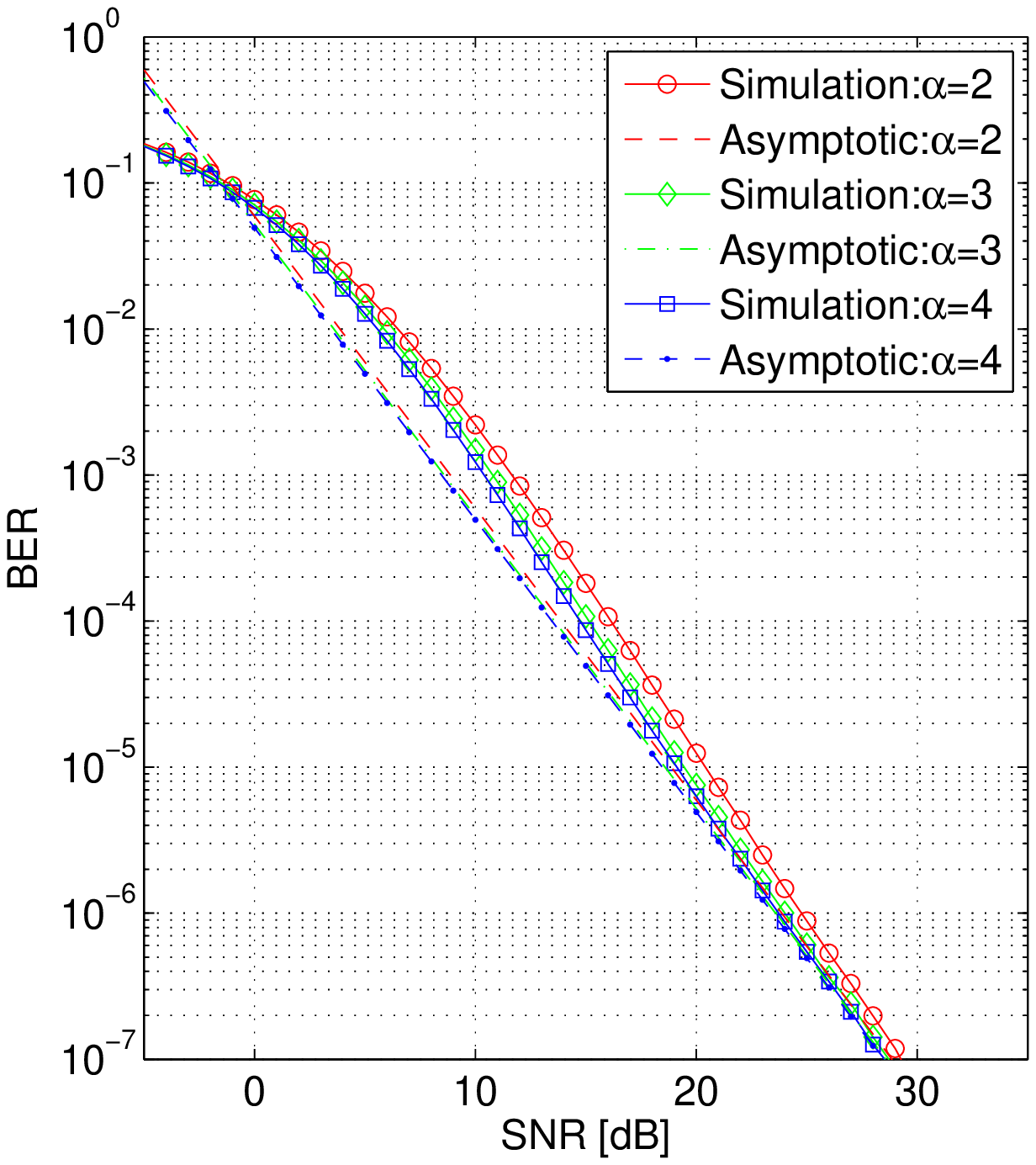}
  }
  \caption{BER of coherent V-BLAST MIMO systems: \(M=2\), \(N=2\).}
  \label{fig:4}
\end{figure*}

In Fig.~\ref{fig:4}(a), we consider a system with \(M=2\), \(N=2\). The Gamma-Gamma turbulence parameters are such that we have a fixed \(\alpha=4\) and a range of \(\beta=1, 1.5, 2, 2.5\) which represent strong (\(\alpha=4\), \(\beta=1\)) to moderate turbulence (\(\alpha=4\), \(\beta=2.5\)) respectively. As seen from Fig.~\ref{fig:4}(a), the diversity gain is 2, which is a counterintuitive result but one that conforms to the behavior of Eqs.~\eqref{equ:32}~and~\eqref{equ:34}. We can also conclude from the results that all the curves where \(\beta\neq 1\) converge to the asymptotic line but that with \(\beta=1\) will not so converge because of Eq.~\eqref{equ:33}. In Eq.~\eqref{equ:33}, the \(P\left( {11 \to 01} \right)\) and \(P\left( {11 \to 10} \right)\) terms will have the same order as \(P\left( {11 \to 00} \right)\), leading to a larger coefficient of the BER. As a result, there is an SNR penalty of approximately 3.5 dB at the Forward Error Correction (FEC) limit of \(10^{-3}\). On the other hand, we also observe that the BER curve will converge to the asymptotic line more quickly with a larger \(\beta\). This phenomenon can also be explained by Eq.~\eqref{equ:33}: the \(P\left( {11 \to 01} \right)\) and \(P\left( {11 \to 10} \right)\) terms will be of higher order and will decay more quickly. Moreover, we can also conclude from Fig.~\ref{fig:4}(a) that the coefficients of the asymptotic lines from Eq.~\eqref{equ:34} are very similar to each other with different values of \(\beta\).

In  Fig.~\ref{fig:4}(b), we consider a system with \(M=2\), \(N=2\) for a range of values of \(\alpha=2, 3, 4\) and fixed \(\beta=1.5\) which represent a strong (\(\alpha=2\), \(\beta=1.5\)) to a moderate turbulence (\(\alpha=4\), \(\beta=1.5\)) respectively. We can conclude that while the coefficients of the asymptotic lines are similar to each other with different \(\alpha\) values, the BER curves will converge more quickly as \(\alpha\) increases.

  \begin{figure}[htb]
  \centering
  \includegraphics[width=4in]{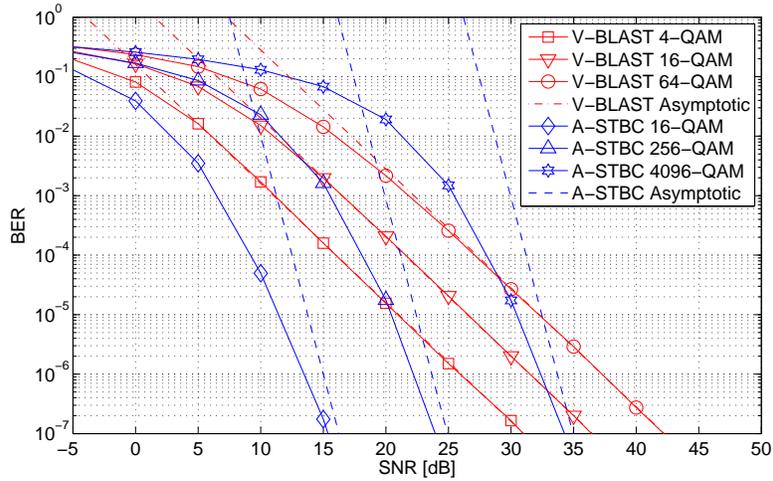}
  \caption{BER comparison of different coding schemes: \(M=2\), \(N=2\), \(\alpha=4\), \(\beta=2\).}
  \label{fig:5}
  \end{figure}
Although the performance of QAM coded MIMO systems presents more difficulty in its analytical solution, it is not impossible to obtain a numerical result. In Fig.~\ref{fig:5}, we compare the BER performance of the V-BLAST scheme with the A-STBC scheme. Both systems have two transmitters and two receivers and the Gamma-Gamma turbulence parameters are \(\alpha=4\), \(\beta=2\), which represent moderate turbulence. In order to achieve the same transmitted bit rate, \(q^M\)-QAMs are applied to the A-STBC scheme while \(q\)-QAMs are applied to the V-BLAST scheme (q=4,16,64). As is seen from Fig.~\ref{fig:5}, although the asymptotic slope of the A-STBC scheme is \(M\beta\) times as large as that of the V-BLAST scheme, the distance between each adjacent asymptotic line of the A-STBC scheme is approximately \(M\) times as great as that of the V-BLAST scheme. This phenomenon implies that the V-BLAST scheme has a better performance in the high SNR regime while the A-STBC scheme is preferable when the SNR is relatively low. 
  
  \begin{figure}[htb]
  \centering
  \includegraphics[width=4in]{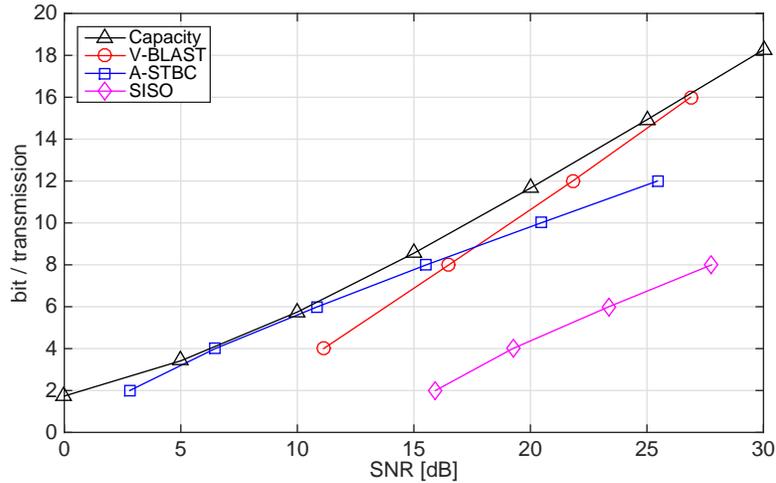}
  \caption{Comparison of different coding schemes: \(M=2\), \(N=2\), \(\alpha=4\), \(\beta=2\).}
  \label{fig:6}
  \end{figure}
In Fig.~\ref{fig:6}, we compare capacities achieved by the V-BLAST scheme and the A-STBC scheme at the FEC limit of \(10^{-3}\). Moreover, 4-QAM to 4096-QAM are applied to match the BER at different SNR values for fair comparison. Both systems have 2 transmitters and 2 receivers and the Gamma-Gamma turbulence parameters are \(\alpha=4\), \(\beta=2\) as used to produce Fig.~\ref{fig:5}. We also present the numerical calculation of the Shannon capacity of such a MIMO Gamma-Gamma channel as well as the corresponding performance of an MLD SISO scheme to provide upper and lower bounds respectively. As is shown in Fig.~\ref{fig:6}, although the performance of the A-STBC system is better in the low SNR area (\(< 18\) dB) attributable to the optimal diversity gain of such a system, V-BLAST is better when the SNR is high, which corresponds to the asymptotic multiplexing performance (the slope of the curve). We can also conclude that the V-BLAST system can achieve the optimal multiplexing gain when comparing it to the capacity limit, which is an existing conclusion in Rayleigh channels \cite{Zheng2003p1073}. So one can conclude from both Figs.~\ref{fig:5}~and~\ref{fig:6} that the V-BLAST system will be preferred in the high SNR area to further improve the transmitted data rate.

\section{Conclusions}
\label{sec:5}

Turbulent optical channel pdfs come in challenging forms resulting in significant difficulties in designing systems that transmit a higher symbol rate than one symbol per transmission in such channels. In this paper, we have proposed a new method to derive the asymptotic BER performance for MIMO systems. Based on the derivations, we have also analyzed the diversity gain of such systems. This process shows that the diversity order will be equal to the number of receivers in the V-BLAST coherent OWC system. Moreover, there is gain on the coefficient rather than on the slope of the BER-SNR curve in the Gamma-Gamma turbulent channel, which is a somewhat counterintuitive result.

By exploiting the spatial multiplexing property of the V-BLAST MIMO regime, we have shown that it is possible to transmit at a higher symbol rate in a Gamma-Gamma turbulent channel with severe turbulence and moderate attenuation. Considering that the spatial multiplexing property is independent of the application of different modulation methods such as PSK and QAM, this regime will be extremely useful when we want to further improve the transmitted data rate in the high SNR area, which is undoubtedly a constantly discussed topic in communication systems.

In the next step, we will investigate the exact BER performance of the coherent V-BLAST MIMO OWC systems. Moreover, we will also consider the trade-off between multiplexing gain and diversity gain in the near future, and thus design a more suitable communication system in certain channel conditions.

\section*{Appendix}

\begin{figure}[htb]
  \centering
  \includegraphics[width=3in]{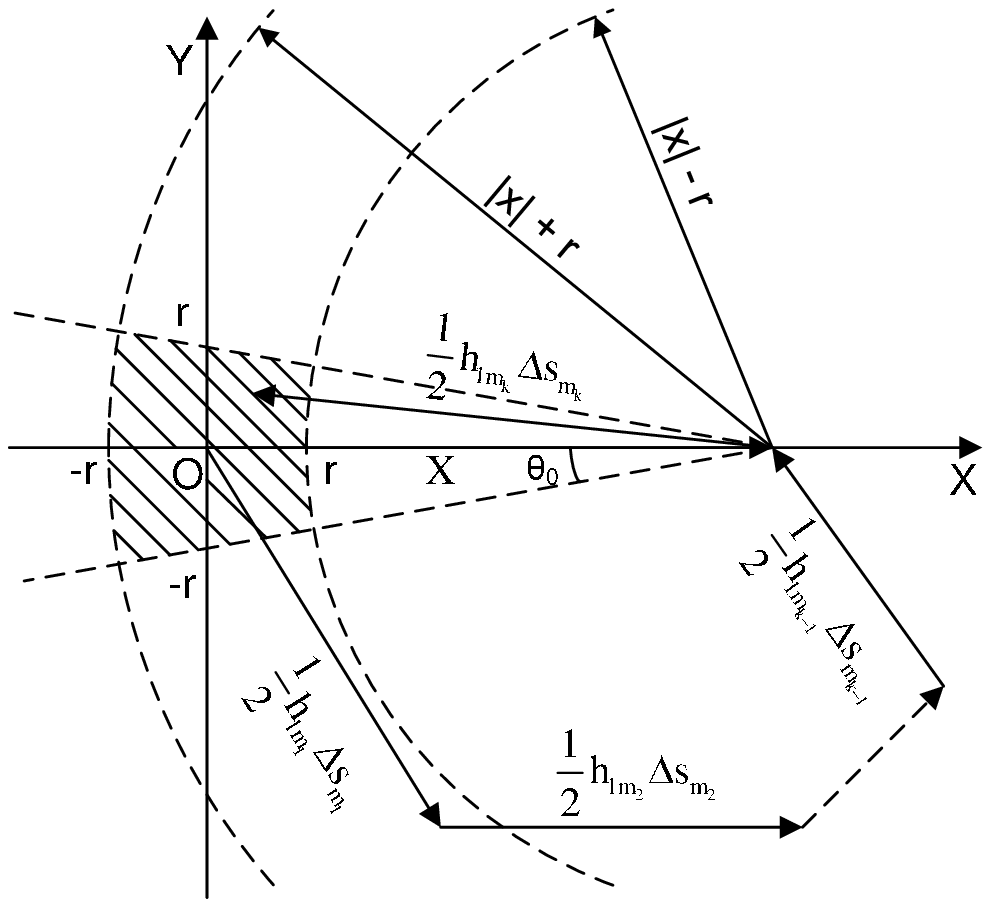}
  \caption{Modified integral area of the MISO system in the complex plane.}
  \label{fig:a1}
  \end{figure}

First of all, we require the cumulative distribution function (cdf) and the probability density function (pdf) of \({\frac{1}{2}\left| {\sum\limits_{1 \le j \le k} {{h_{1{m_j}}}\Delta {s_{{m_j}}}} } \right|} < r\) where \(r \le r_x\). Without loss of generality, we can put \({X=\frac{1}{2} {\sum\limits_{1 \le j \le k-1} {{h_{1{m_j}}}\Delta {s_{{m_j}}}} }}\) on the \(x\)-axis because it is a circular symmetric variable. However, the calculation of the probability of \({\frac{1}{2}\left| {\sum\limits_{1 \le j \le k} {{h_{1{m_j}}}\Delta {s_{{m_j}}}} } \right|} < r\) is not tractable when the circular area is considered and so here we develop an approximation. This is based on a modified scenario shown in Fig.~\ref{fig:a1}.
Furthermore, in the high SNR regime, the shaded region in Fig.~\ref{fig:a1} will become approximately square with sides of length \(2r\). Moreover, the conditional probability \(P\left\{ {\left. {\left| {\frac{1}{2}\left| {{h_{1{m_k}}}\Delta {s_{{m_k}}}} \right| - \left| X \right|} \right| < r} \right|{\bf{I}}} \right\}\) with arbitrary \({\bf{I}} = \left[ {{I_{1{m_1}}},{I_{1{m_2}}}, \ldots ,{I_{1{m_{k - 1}}}}} \right]\) will become approximately a constant in the small area. Since the area of the square is \(4r^2\) while that of the circle in Fig.~\ref{fig:2} is \(\pi r^2\), we can conclude that the probability of Fig.~\ref{fig:a1} is \(4/\pi\) times as large as the probability of Fig.~\ref{fig:2}. In summary, we can first integrate the shaded region in  Fig.~\ref{fig:a1} and then simply multiply it by \(\pi / 4\) to obtain the probability of Fig.~\ref{fig:2}.

To calculate the probability of \({\frac{1}{2}\left| {\sum\limits_{1 \le j \le k} {{h_{1{m_j}}}\Delta {s_{{m_j}}}} } \right|} \) lying in the shaded region in Fig.~\ref{fig:a1}, we can write it in the form of an integral thus:
  \begin{equation}\label{equ:a16}
  {P_{rect}} \buildrel \Delta \over = \int\limits_{\bf{I}} {P\left\{ {\left. {\left| {\arg \left( {{h_{1{m_k}}}\Delta {s_{{m_k}}}} \right)} \right| < {\theta _0}} \right|{\bf{I}}} \right\}P\left\{ {\left. {\left| {\frac{1}{2}\left| {{h_{1{m_k}}}\Delta {s_{{m_k}}}} \right| - \left| X \right|} \right| < r} \right|{\bf{I}}} \right\}{f_{\bf{I}}}\left( {\bf{I}} \right){\rm{d}}{\bf{I}}},
  \end{equation}
where \({\bf{I}} = \left[ {{I_{1{m_1}}},{I_{1{m_2}}}, \ldots ,{I_{1{m_{k - 1}}}}} \right]\) is the normalized irradiance of the channel from the first transmitter to the first receiver. Considering
  \begin{equation}\label{equ:a17}
  P\left\{ {\left. {\left| {\arg \left( {{h_{1{m_k}}}\Delta {s_{{m_k}}}} \right)} \right| < {\theta _0}} \right|{\bf{I}}} \right\} = \frac{{2r}}{{2\pi \left| X \right|}},
  \end{equation}
and \(r\) will be infinitesimally small so that \(f_I \left( {\left| X \right|}^2 \right) \) is approximately constant over the range of the integral so:

  \begin{equation}\label{equ:a18}
  \begin{aligned}
  P\left\{ {\left. {\left| {\frac{1}{2}\left| {{h_{1{m_k}}}\Delta {s_{{m_k}}}} \right| - \left| X \right|} \right| < r} \right|{\bf{I}}} \right\} &= \int_{{{2\left( {\left| X \right| - r} \right)} \mathord{\left/
   {\vphantom {{2\left( {\left| X \right| - r} \right)} {\left| {\Delta {s_{{m_k}}}} \right|}}} \right.
 \kern-\nulldelimiterspace} {\left| {\Delta {s_{{m_k}}}} \right|}}}^{{{2\left( {\left| X \right| + r} \right)} \mathord{\left/
 {\vphantom {{2\left( {\left| X \right| + r} \right)} {\left| {\Delta {s_{{m_k}}}} \right|}}} \right.
 \kern-\nulldelimiterspace} {\left| {\Delta {s_{{m_k}}}} \right|}}} {{f_h}\left( {{h_{1{m_k}}}} \right)} {\rm{d}}{h_{1{m_k}}}\\
 &= \int_{4{{{{\left( {\left| X \right| - r} \right)}^2}} \mathord{\left/
 {\vphantom {{{{\left( {\left| X \right| - r} \right)}^2}} {{{\left| {\Delta {s_{{m_k}}}} \right|}^2}}}} \right.
 \kern-\nulldelimiterspace} {{{\left| {\Delta {s_{{m_k}}}} \right|}^2}}}}^{{{4{{\left( {\left| X \right| + r} \right)}^2}} \mathord{\left/
 {\vphantom {{4{{\left( {\left| X \right| + r} \right)}^2}} {{{\left| {\Delta {s_{{m_k}}}} \right|}^2}}}} \right.
 \kern-\nulldelimiterspace} {{{\left| {\Delta {s_{{m_k}}}} \right|}^2}}}} {{f_I}\left( I \right)} {\rm{d}}I\\
 &= \frac{{16\left| X \right|r}}{{{{\left| {\Delta {s_{{m_k}}}} \right|}^2}}}{f_I}\left( {\frac{{4{{\left| X \right|}^2}}}{{{{\left| {\Delta {s_{{m_k}}}} \right|}^2}}}} \right),
  \end{aligned}
  \end{equation}
  
we get the form of Eq.~\eqref{equ:a16} as:
  \begin{equation}\label{equ:a19}
  {P_{rect}} = \int\limits_{\bf{I}} {\frac{{2r}}{{2\pi \left| X \right|}} \cdot \frac{{16\left| X \right|r}}{{{{\left| {\Delta {s_{{m_k}}}} \right|}^2}}}{f_I}\left( {\frac{{4{{\left| X \right|}^2}}}{{{{\left| {\Delta {s_{{m_k}}}} \right|}^2}}}} \right){f_{\bf{I}}}\left( {\bf{I}} \right){\rm{d}}{\bf{I}}}  = \int\limits_{\bf{I}} {\frac{{16{r^2}}}{{\pi {{\left| {\Delta {s_{{m_k}}}} \right|}^2}}}{f_I}\left( {\frac{{4{{\left| X \right|}^2}}}{{{{\left| {\Delta {s_{{m_k}}}} \right|}^2}}}} \right){f_{\bf{I}}}\left( {\bf{I}} \right){\rm{d}}{\bf{I}}}.
  \end{equation}

Considering \(P_{rect}=4/\pi\cdot P_{circle}\), we arrive at the cdf of \( {\left| {{h_{11}} + {h_{12}}} \right| < r}\) thus:
  \begin{equation}\label{equ:a21}
  {F_1}\left( r \right) = {P_{circle}} = \frac{\pi }{4}{P_{rect}} = {C_r}{r^2},
  \end{equation}
where
  \begin{equation}\label{equ:a22}
  {C_r} = \int\limits_{\bf{I}} {\frac{4}{{{{\left| {\Delta {s_{{m_k}}}} \right|}^2}}}{f_I}\left( {\frac{{4{{\left| X \right|}^2}}}{{{{\left| {\Delta {s_{{m_k}}}} \right|}^2}}}} \right){f_{\bf{I}}}\left( {\bf{I}} \right){\rm{d}}{\bf{I}}}.
  \end{equation}

Moreover, we can also conclude that the pdf can be written as:
  \begin{equation}\label{equ:a23}
{f_1}\left( r \right) = \frac{{{\rm{d}}{F_1}\left( r \right)}}{{{\rm{d}}r}} = 2{C_r}r.
  \end{equation}

\section*{Funding}

China Scholarship Council (CSC) (201706070081); University of Warwick (1754694).

\bibliography{paperbib}

\end{document}